\documentclass[a4paper,fleqn,usenatbib]{mnras}

\usepackage{newtxtext,newtxmath}

\usepackage[T1]{fontenc}
\usepackage{ae,aecompl}

\usepackage[dvipsnames]{xcolor}
\usepackage{float}

\usepackage{graphicx}	
\usepackage{amsmath}	
\usepackage{amssymb}	

\title[Tidal origin of NGC~1427A]{Tidal origin of NGC~1427A in the Fornax cluster}

\author[K. Lee-Waddell et al.]{
K. Lee-Waddell$^{1}$\thanks{E-mail: karen.lee-waddell@csiro.au},
P. Serra$^{2,1}$, B. Koribalski$^{1}$, A. Venhola$^{3,4}$, E. Iodice$^{5}$, B. Catinella$^{6}$, 
\newauthor 
\hspace{0.5mm} L. Cortese$^{6}$, R. Peletier$^{3}$, A. Popping$^{6,7}$, O. Keenan$^{8}$, M. Capaccioli$^{9}$
\\
$^{1}$CSIRO Astronomy and Space Sciences, Australia Telescope National Facility, PO Box 76, Epping, NSW 1710, Australia\\
$^{2}$INAF -- Osservatorio Astronomico di Cagliari, Via della Scienza 5, I-09047 Selargius (CA), Italy\\
$^{3}$Kapteyn Astronomical Institute, University of Groningen, PO Box 800, NL-9700 AV Groningen, the Netherlands\\
$^{4}$Astronomy Research Unit, University of Oulu, FI-90014, Finland\\
$^{5}$INAF -- Astronomical Observatory of Capodimonte, via Moiariello 16, Naples, I-80131, Italy\\
$^{6}$International Centre for Radio Astronomy Research, The University of Western Australia, 35 Stirling Hwy, Crawley, WA 6009, Australia\\
$^{7}$CAASTRO: ARC Centre of Excellence for All-sky Astrophysics, Australia\\
$^{8}$School of Physics and Astronomy, Cardiff University, Queens Buildings, The Parade, Cardiff CF24 3AA, United Kingsdom\\
$^{9}$Dip.di Fisica Ettore Pancini, University of Naples ``Federico II,'' C.U. Monte SantAngelo, Via Cinthia, I-80126, Naples, Italy\\
}

\date{Accepted 2017 October 26. Received 2017 October 15; in original form 2017 March 31}

\pubyear{2017}

\begin{document}
\label{firstpage}
\pagerange{\pageref{firstpage}--\pageref{lastpage}}
\maketitle

\begin{abstract} 
We present new H{\sc{i}} observations from the Australia Telescope Compact Array and deep optical imaging from OmegaCam on the VLT Survey Telescope of NGC~1427A, an arrow-shaped dwarf irregular galaxy located in the Fornax cluster.  The data reveal a star-less H{\sc{i}} tail that contains $\sim$10\% of the atomic gas of NGC~1427A as well as extended stellar emission that shed new light on the recent history of this galaxy.  Rather than being the result of ram pressure induced star-formation, as previously suggested in the literature, the disturbed optical appearance of NGC~1427A has tidal origins.  The galaxy itself likely consists of two individual objects in an advanced stage of merging.  The H{\sc{i}} tail may be made of gas expelled to large radii during the same tidal interaction.  It is possible that some of this gas is subject to ram pressure, which would be considered a secondary effect and imply a northwest trajectory of NGC 1427A within the Fornax cluster.

\end{abstract}




\section{Introduction}
\label{sec:intro}

\vspace{5mm}

Galaxy clusters are often turbulent environments that host various types of galaxies at different stages of evolution (e.g. \citealt{mih2005}, \citealt{tol2014}).  Tidal interactions in the outskirts give way to ram pressure stripping in the central regions of clusters (\citealt{too1972}, \citealt{gun1972}), offering the ability for side-by-side comparison of these two mechanisms.  

In low-density regions, encounters between galaxies can produce extended tidal tails of stars and gas.  These tails can constrain the properties of interaction event and be used to investigate the environmental dynamics of a system (e.g.~\citealt{bou2007}).  As galaxies move towards the cluster potential, the gas in their outer disc and/or in any of these tidal tails becomes susceptible to ram pressure stripping.  The interior region of clusters is typically dominated -- in number -- by dwarf elliptical galaxies (dEs; \citealt{tol2014}).  Many of these dEs started as star-forming irregular galaxies and experienced rapid evolution as the intracluster medium (ICM) depleted their gas content through ram pressure stripping (e.g.~\citealt{ken2014}).  In this paper we study the properties of NGC~1427A, a dwarf galaxy evolving near (possibly within) the Fornax cluster.

\subsection{The Fornax dwarf NGC~1427A}
\label{sec:intro2}

The Fornax cluster is a compact (virial radius of 0.7 Mpc), low mass ($7 \times 10^{13}$ $M_{\odot}$ within a 1.4 Mpc radius) cluster consisting of 108 spectroscopically confirmed galaxies -- with magnitudes between -16 < $M_B$ < -13.5, a mean velocity of 1493 $\pm$ 36 km s$^{-1}$, and a velocity dispersion of 374 $\pm$ 26 km s$^{-1}$ -- located $\sim$20 Mpc away \citep{dri2001a}.  Its high central galactic density and low velocity dispersion \citep{dri2001b}, suggest that tidal interactions could be playing a prominent role in the evolution of cluster members, in addition to ram pressure stripping.  NGC~1399 is considered to be the central galaxy of the cluster and located within a projected distance of $\sim$100 kpc from this centre is NGC~1427A, a bright arrow-shaped dwarf irregular (dIrr).

One of the earliest detailed optical studies of NGC~1427A was conducted by \citet{cel1997} who attributed the distorted shape of this galaxy to tidal forces and suggested that a northern stellar clump might be a separate object that recently interacted with NGC~1427A, causing a burst in star-formation.  B,V, I and H$\alpha$ imaging by \citet{hil1997} revealed that the majority of the brighter OB associations and H{\sc{ii}} regions in NGC~1427A are aligned along its southwest edge.  They identified $\sim$30 stellar clusters -- uniformly distributed over the entire galaxy -- that appear to have mean ages less than 2 Gyr, indicating recent starbursts as the result of either an interacting interloper or ram pressure triggered star-formation \citep{hil1997}.  \citet{cha2000} used long-slit spectroscopy on the seven brightest H{\sc{ii}} regions across NGC~1427A to show that the aforementioned northern clump has the same velocity pattern as the rest of the galaxy and is likely a part of NGC~1427A (rather than an intruder).  With deep optical imaging, \citet{mor2015} estimate that the most recent episode of star-formation in NGC~1427A began $\sim$4 Myr ago and that the current star cluster formation rate of this dIrr is consistent with other starburst galaxies.  Accordingly, the morphological properties of NGC~1427A were deemed the result of its passage through and interaction with the hot intracluster gas of the cluster (\citealt{cha2000}; \citealt{mor2015}).  

In this paper, we use new radio and optical observations to investigate whether ram pressure is the primary hydrodynamic interaction that is affecting NGC~1427A.  We present arcminute-resolution H{\sc{i}} observations -- taken with the Australia Telescope Compact Array (ATCA) -- and deep optical imaging from OmegaCam on the VLT Survey Telescope (VST) that elucidate the structure of and evolutionary history for NGC~1427A.  In Section \ref{sec:obs}, we describe the observations of the Fornax cluster with focus on NGC~1427A.  Section \ref{sec:results} summarizes the measured results and describes the newly resolved H{\sc{i}} tail and key stellar features of NGC~1427A.  In Section \ref{sec:discuss}, we compare our data with results and conclusions presented in the literature as well as discuss the possible origin of this tail.  We present our final conclusions in Section \ref{sec:conclude}.

\section{Observations}
\label{sec:obs}

There have been numerous surveys of the Fornax cluster (e.g. \citealt{bur1996}, \citealt{dri2001a}, \citealt{wau2002}).  Many of these surveys are broadly focused on general detection and cluster properties as a whole.  For a more in-depth study of NGC~1427A, we utilized a small portion of the data from two fairly recent high-resolution surveys, as described in this section.

\subsection{ATCA H{\sc{i}} observations}
A blind H{\sc{i}} survey of the Fornax cluster was conducted using ATCA's 750B array in late 2013 (project code C2894; PI: P. Serra).  A 13 deg$^2$ field centred on the cluster was observed using multiple pointings over a 28 day period (totalling 331 hr).  The fully mosaicked field comprises a hexagonal grid of 756 pointings with 8.7 arcmin spacing ($\sim$2 times better than Nyquist sampling at 1.4 GHz). Each day (i.e.~during each $\sim$12 hr observation) we mosaicked a field of $\sim$$0.7 \times 0.7$ deg$^2$ using 27 pointings. We cycled through the 27 pointings revisiting each of them every 800 seconds.  Although this method does result in a small loss of uv coverage for the longest baseline of interest (750 m), it produces an acceptable slew-time overhead of 10 percent the total observing time.  Observations of PKS B1934-638 were used for flux and bandpass calibrations.  PKS 0332-403 was observed at 1.5 hr intervals (between on-source scans) for phase calibrations.  The 64 MHz bandwidth, centred at 1396 MHz, was divided into 2048 channels providing a channel resolution of 31.25 kHz or 6.6 km s$^{-1}$.

\begin{figure*}
\begin{center}
  \includegraphics[width=165mm,trim={25mm 29mm 25mm 34.3mm},clip]{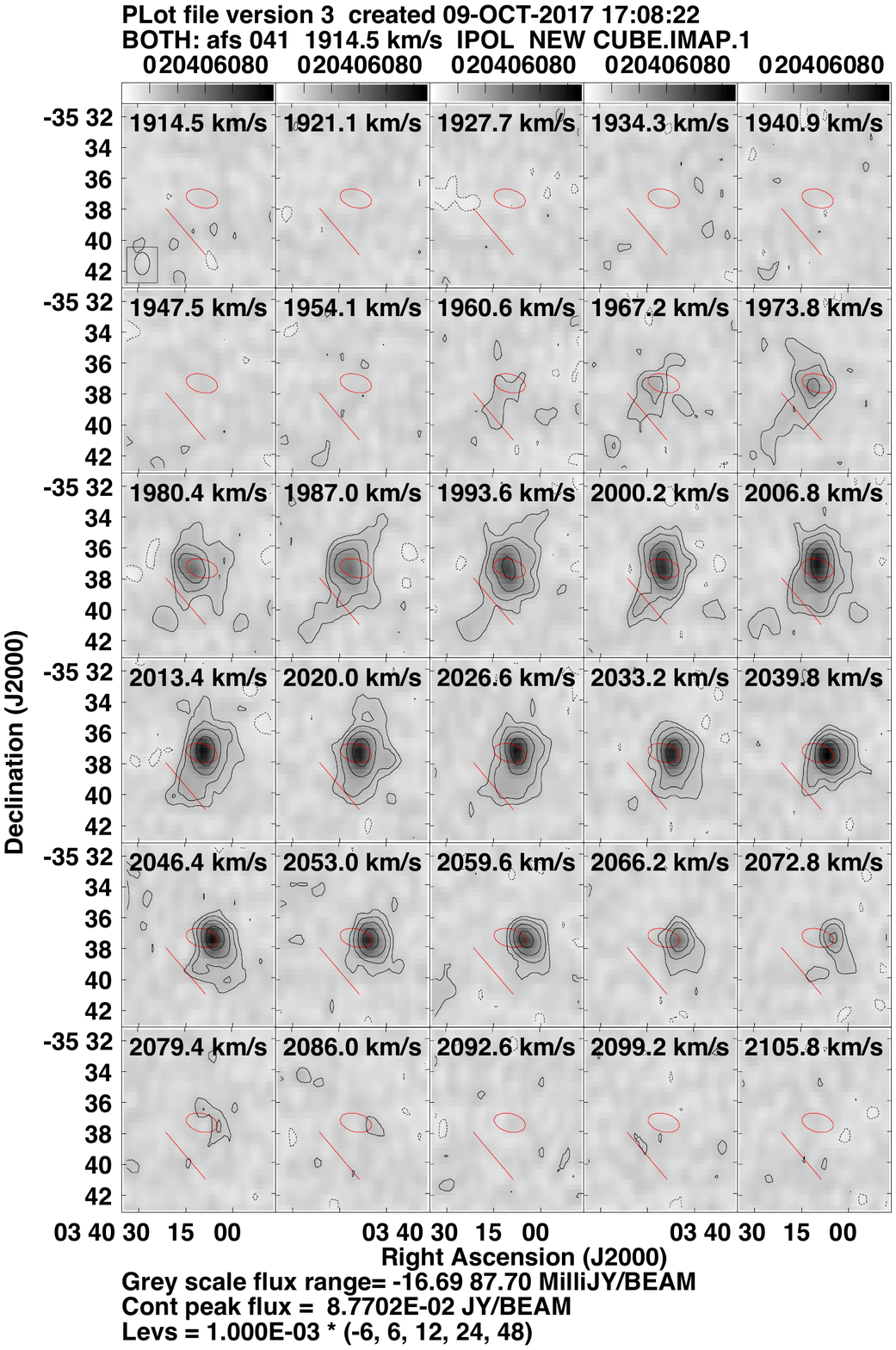}
  \caption{ATCA H{\sc{i}} channel maps of NGC~1427A.  The velocity of each channel is indicated at the top of each panel and the shape of synthesized beam is shown in the bottom left of the first panel.  The H{\sc{i}} intensity contours are at (-6, 6, 12, 24, 48) mJy beam$^{-1}$, with the negative contour indicated as the dashed line.  The red ellipses represent the star-forming optical shape of the galaxy (see Fig.~\ref{fig:ugri}) and the red lines show the approximate boundary between H{\sc{i}} in the main body of NGC~1427A and H{\sc{i}} in the tail (see text and Table~\ref{table:HI} for further details). \vspace{20mm}
\label{fig:chan}}
\end{center}
\end{figure*}

The data were reduced and imaged using the {\sc{miriad}} software package \citep{sau1995}.  After flagging and calibration, the individual pointings were mosaicked to increase the signal-to-noise of the H{\sc{i}} sources for better cleaning during the imaging process.  We subtracted the radio continuum emission from the visibilities by fitting a polynomial to the line-free channels (using the {\sc{uvlin}} task in {\sc{miriad}}).  For each pointing, the order of the polynomial was set to the lowest value that resulted in no residual continuum emission in the H{\sc{i}} cube (typically 3).  The naturally weighted H{\sc{i}} sub-cube of the region around NGC~1427A has an 86 $\times$ 59 arcsec synthesized beam and an RMS noise of 3.1 mJy beam$^{-1}$.  Fig.~\ref{fig:chan} shows the channel maps of the H{\sc{i}} associated with NGC~1427A.

\subsection{VST data}

The optical observations of NGC~1427A are part of the Fornax Deep Survey (FDS) with OmegaCam on the VST (\citealt{iod2016}; \citealt{dab2016}; \citealt{iod2017}; \citealt{ven2017}).  OmegaCam has a 1 deg $\times$ 1 deg field of view and comprises an array of 8 $\times$ 4 CCDs, each with 2144 $\times$ 4200 pixels.  OmegaCam's unbinned pixel size of 0.21 arcsec provides a well-sampled point spread function for the observations, which have an average FWHM of $\sim$1 arcsec.

The currently available data were taken during visitor mode runs in November 2013, November 2014 and November 2015 (ESO P92, P94 and P96 respectively).  The VST mosaic of the inner two square degrees around the core of the Fornax cluster is presented by \citet[see also https://www.eso.org/public/unitedkingdom/news/eso1612/?lang]{iod2016}.  In this paper we used a small snapshot around NGC1427A taken from the whole mosaic in the four OmegaCam $u'$, $g'$, $r'$ and $i'$-bands, kindly provided by the FDS PIs.  The observing strategy and datasets are described by \citet{iod2016}.  

\begin{figure*}
\begin{center}
  \includegraphics[width=140mm]{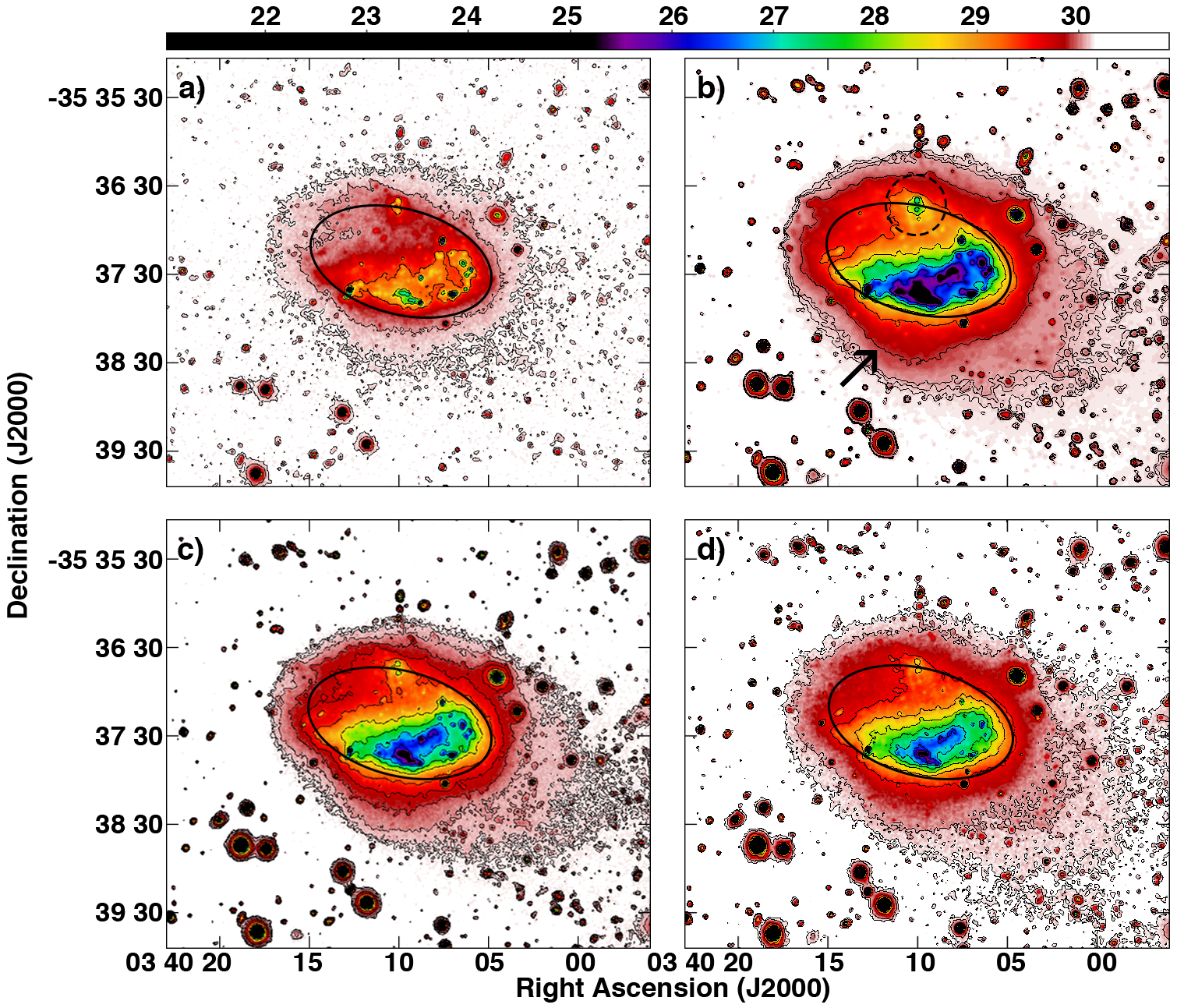}
  \caption{Smoothed VST images of NGC~1427A.  The black ellipses are the same as the red ones shown in Fig.~\ref{fig:chan}.  a) $u'$-band. b) $g'$-band. The dashed circle indicates the northern stellar clump and the arrow points to a very faint stellar over-density `bump' (see text for further details).  c) $r'$-band. d) $i'$-band.
\label{fig:ugri}}
\end{center}
\end{figure*}

The OmegaCam images presented in this paper were processed by using the {\sc{AstroWISE}} pipeline, described in detailed by \citet[see also \citealt{mcf2013}]{ven2017}.  Instrumental corrections include the removal of bias and scattered light (background). The images have also been corrected for uneven illumination -- by applying flat-field and illumination corrections -- and calibrated to have a zero point of 0.  These observations have a 1$\sigma$ photometric accuracy of 0.04 mag, 0.02 mag, 0.02 mag and 0.03 mag for $u'$, $g'$, $r'$ and $i'$, respectively (please refer to \citealt{ven2017} for further details).  Fig.~\ref{fig:ugri} shows optical images of NGC~1427A in each band.  In order to highlight the low surface brightness structures in these images, we performed a Gaussian smoothing with a radius of 10 pixels.

\section{Results}
\label{sec:results}

\subsection{ATCA H{\sc{i}} results}

The ATCA observations provide sufficient resolution to resolve the extended H{\sc{i}} structure of NGC~1427A (see Fig.~\ref{fig:chan}).  The properties of this galaxy as a whole (i.e.~H{\sc{i}} total flux, $F_{H_I}$, central velocity, $v_{H_I}$, and velocity widths, $W_{50}$ and $W_{20}$, at 50\% and 20\% of its respective peak flux), which are presented in Table~\ref{table:HI}, were measured using the H{\sc{i}} source finding application {\sc{SoFiA}} \citep{ser2015} and verified using manual measurement techniques (i.e.~{\sc{mbspect}} task in {\sc{miriad}} and the spectral profile plotting tool in {\sc{casa}}; \citealt{mcm2007}).  The difference in $F_{H_I}$ obtained by each measurement method was added in quadrature to the noise in the cube resulting in a $\sim$10\% uncertainty.  The global measurements of NGC~1427A are consistent with values in the literature measured by \citep{bur1996} and \citep{kor2004} using the single-dish Parkes radio telescope and presented in Table~\ref{table:HI_parkes}.

\begin{table}
 \centering
 \begin{minipage}{85mm}
 \caption{H{\sc{i}} properties of NGC~1427A, measured by ATCA.  The uncertainties in $F_{H_I}$ and $M_{H_I}$ for the H{\sc{i}} tail reflect the uncertainty in separating this component from the H{\sc{i}} core region.}
 \label{table:HI}
\begin{tabular}{ l c c c}  
\hline
							& NGC~1427A		& H{\sc{i}} tail			& H{\sc{i}} core region\\ 
\hline 
Peak flux (Jy)					& 0.27 $\pm$ 0.02	& 0.038 $\pm$ 0.003		& 0.24 $\pm$ 0.02\\
$F_{H_I}$ (Jy km s$^{-1}$)		& 22 $\pm$ 2		& 2.1 $\pm$ 0.6		& 20 $\pm$ 2\\
$v_{H_I}$ (km s$^{-1}$)			& 2021 $\pm$ 3	& 1997 $\pm$ 3		& 2018 $\pm$ 3\\
$W_{50}$ (km s$^{-1}$)			& 80 $\pm$ 3		& 68 $\pm$ 3			& 77 $\pm$ 3\\
$W_{20}$ (km s$^{-1}$)			& 110 $\pm$ 3		& 76 $\pm$ 3			& 107 $\pm$ 3\\
$M_{H_I}$ ($10^9$ $M_{\odot}$)	& 2.1 $\pm$ 0.2	& 0.20 $\pm$ 0.06 		& 1.9 $\pm$ 0.2\\
\hline
\end{tabular} 
\end{minipage}
\end{table}

\begin{table}
 \centering
 \begin{minipage}{85mm}
 \caption{H{\sc{i}} properties of NGC~1427A, measured by Parkes}
 \label{table:HI_parkes}
\begin{tabular}{ l c c}  
\hline
						&\citet{bur1996}		&\citet{kor2004}\\ 
\hline 
Peak flux (Jy)				& --					& 0.261 $\pm$ 0.019 \\
$F_{H_I}$ (Jy km s$^{-1}$)	& 23.1 $\pm$ 1.2		& 22.5 $\pm$ 3.0 \\
$v_{H_I}$ (km s$^{-1}$)		& 2027.8 $\pm$ 0.8		& 2029 $\pm$ 4\\
$W_{50}$ (km s$^{-1}$)		& 83.4 $\pm$ 1.5		& 86\\
$W_{20}$ (km s$^{-1}$)		& 118.8 $\pm$ 1.0		& 119\\
\hline
\end{tabular} 
\end{minipage}
\end{table}

\begin{figure*}
\begin{center}
  \includegraphics[width=178mm]{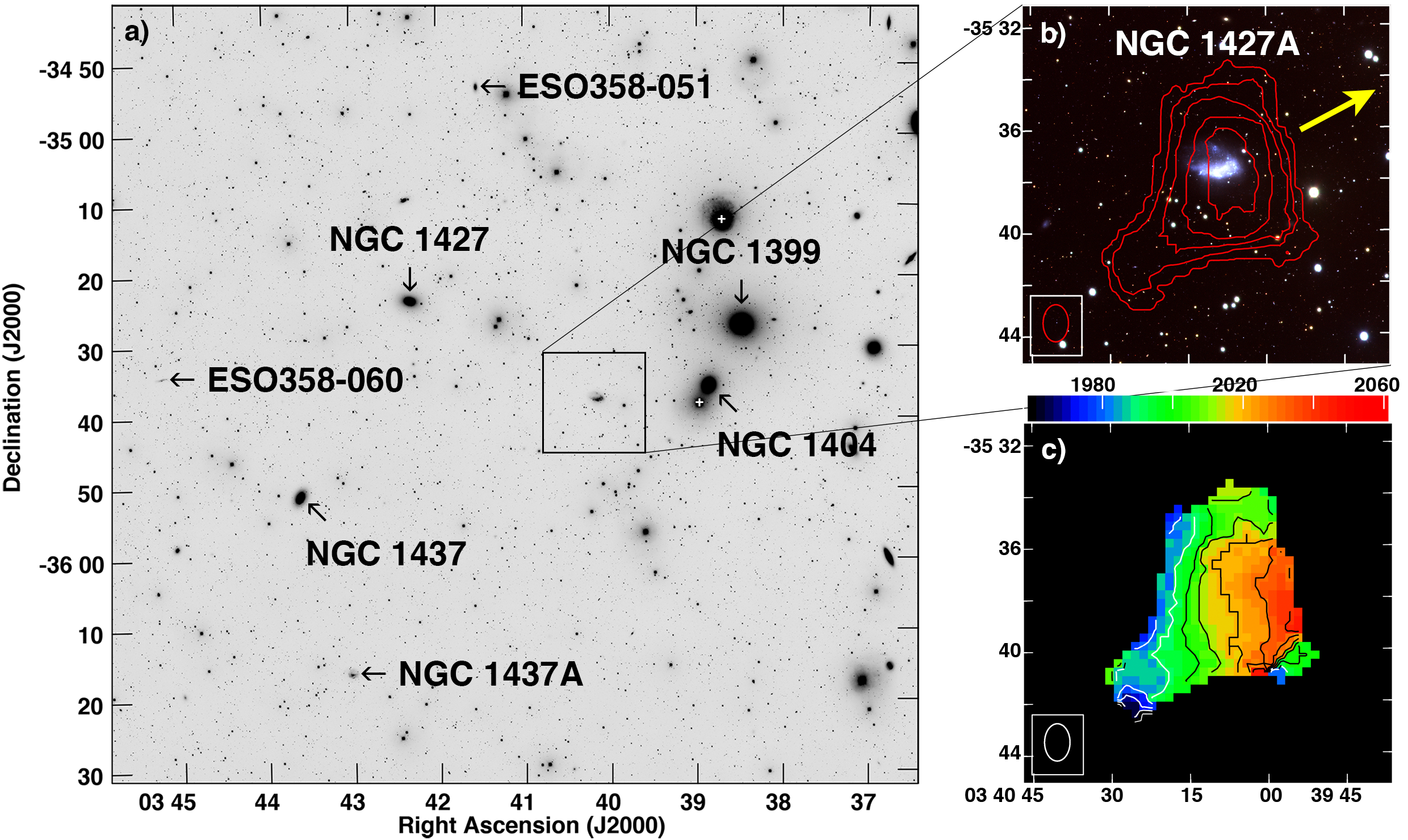}
  \caption{a) VST $r'$-band image of central region of the Fornax cluster.  Prominent cluster member are indicated by name and bright foreground stars are denoted by the white crosses.  b) ATCA total intensity H{\sc{i}} (moment-0) contours from the {\sc{SoFiA}} masked cube superimposed on a VST $g'r'i'$-band composite image of NGC~1427A.  Contours are at (0.5, 1, 2, 5, 10) $\times$ $10^{20}$ atoms cm$^{-2}$.  The yellow arrow points towards the central galaxy, NGC~1399.  c) ATCA H{\sc{i}} velocity (moment-1) map.  Contours are shown at 10 km s$^{-1}$ increments.
\label{fig:central}}
\end{center}
\end{figure*}

Fig.~\ref{fig:central} shows the total intensity (i.e. moment-0) H{\sc{i}} contours superimposed on a 3-colour optical image of NGC~1427A, its moment-1 velocity map and the spatial location of this galaxy with respect to other Fornax cluster members.  Within the ATCA cube, the only other H{\sc{i}} features within an $\sim$1 deg ($\sim$350 kpc) and $\sim$1000 km s$^{-1}$ radius of NGC~1427A are ESO358-051, NGC~1437A, and ESO 358-60, which were previously characterized by HIPASS, as well as a new H{\sc{i}} detection, NGC~1437 (see Section \ref{sec:discuss} for further details about these neighbouring galaxies). 

The H{\sc{i}} channel maps in Fig.~\ref{fig:chan} show a bright central `core' and up to three gaseous extensions.  The most prominent extension -- also clearly visible in Fig.~\ref{fig:central}b,c -- is a newly resolved H{\sc{i}}-rich tail extending towards the southeast of the optical centre of the galaxy.  There also appears to be fainter H{\sc{i}} extensions to the north and southwest of the core.  Manually fitted ellipses were used to extract spectral profiles and derive the H{\sc{i}} properties of the H{\sc{i}} tail and the core region (the latter comprises all contiguous emission not included in the former).  In Fig.~\ref{fig:chan} we indicate the approximate measurement boundary between the tail and core region.  Shifting this line closer to the core ellipse can easily add up to 30\% more H{\sc{i}} flux (and mass) to the tail.  We describe the various components of NGC~1427A in the following subsections.

\subsubsection{H{\sc{i}} tail}
\label{subsec:HI_tail}

The H{\sc{i}} tail of NGC~1427A spans a velocity width of $W_{50} = 68 \pm 3$ km s$^{-1}$ centred at 1997 km s$^{-1}$.  With $F_{H_I}$ = 2.1 Jy km s$^{-1}$, this gaseous tail contains about 10\% of the H{\sc{i}} mass ($M_{H_I}$) of NGC~1427A.  Considering that the radius of the central H{\sc{i}} core of NGC~1427A (measured parallel to the tail) is $\sim$15 kpc, then the tail extends $>$20 kpc beyond the H{\sc{i}} core (or $>$25 kpc beyond the stellar core).  The alignment of this H{\sc{i}} tail has significant implications on the direction of any ram pressure forces acting on NGC~1427A, which will be discussed in Section~\ref{sec:discuss}.

\subsubsection{H{\sc{i}} core region}
\label{subsec:HI_core}

The brightest H{\sc{i}} contours in Fig.~\ref{fig:chan} coincide with the star-forming stellar core, indicated by the ellipses, of NGC~1427A.  The distribution of most of the H{\sc{i}} in the core is relatively symmetric and shows signs of rotation -- with a clear velocity gradient -- across the major axis of the disk (see also Fig.~\ref{fig:central}c).  At the 2$\sigma$-level, there is possibly an H{\sc{i}} `counter-tail' to the north of core at $\sim$1990 km s$^{-1}$ and a faint H{\sc{i}} cloud to southwest of the stellar core at $\sim$2003 km s$^{-1}$.  These two marginally detected features appear to only span two velocity channels ($\sim$10 km s$^{-1}$); nevertheless, their coincidence with other features warrants some mention.  The possible counter-tail appears to be directly across from main H{\sc{i}} tail at 1987-1994 km s$^{-1}$ in Fig.~\ref{fig:chan}, which is consistent with tidally-formed features (see \citealt{too1972}).  Whereas, the faint H{\sc{i}} cloud spatially coincides with some extended stellar emission (see Section~\ref{subsec:sw_optical}).  

\subsection{VST results}

The global optical properties of this dIrr were measured from the original, unsmoothed VST images -- using {\sc{SExtractor}} \citep{ber1996}, with the $r'$-band image as the reference for data extraction -- and are presented in Table~\ref{table:opt}.  $R_{50}$ and $R_{90}$ indicate the effective radii containing 50\% and 90\% of the light.  Each listed colour has been computed from the extracted apparent magnitudes ($m_{\lambda}$) in a Kron-like elliptical aperture for all bands.  The stellar mass ($M_*$) of NGC~1427A has been estimated using the empirical relation calibrated by \citet{tay2011} and the assumed Fornax cluster distance, of 20 Mpc \citep{dri2001a}, to the source.

\begin{table}
 \centering
 \begin{minipage}{40mm}
 \caption{Global optical properties of NGC~1427A}
 \label{table:opt}
\begin{tabular}{ l c c}
\hline
Property	& Value	& Units\\ 
\hline 
$R_{50}$	& 34.2	& arcsec\\
$R_{90}$	& 67.3	& arcsec\\
$m_{r'}$	& 12.86	& mag\\
$u'-r'$ 	& 1.27	& mag\\
$g'-r'$	& 0.31	& mag\\
$r'-i'$	& 0.20	& mag\\
$M_* $	& $1.1 \times 10^{9}$	& $M_{\odot}$\\
\hline
\end{tabular} 
\end{minipage}
\end{table}

\begin{figure}
\begin{center}
  \includegraphics[width=84mm]{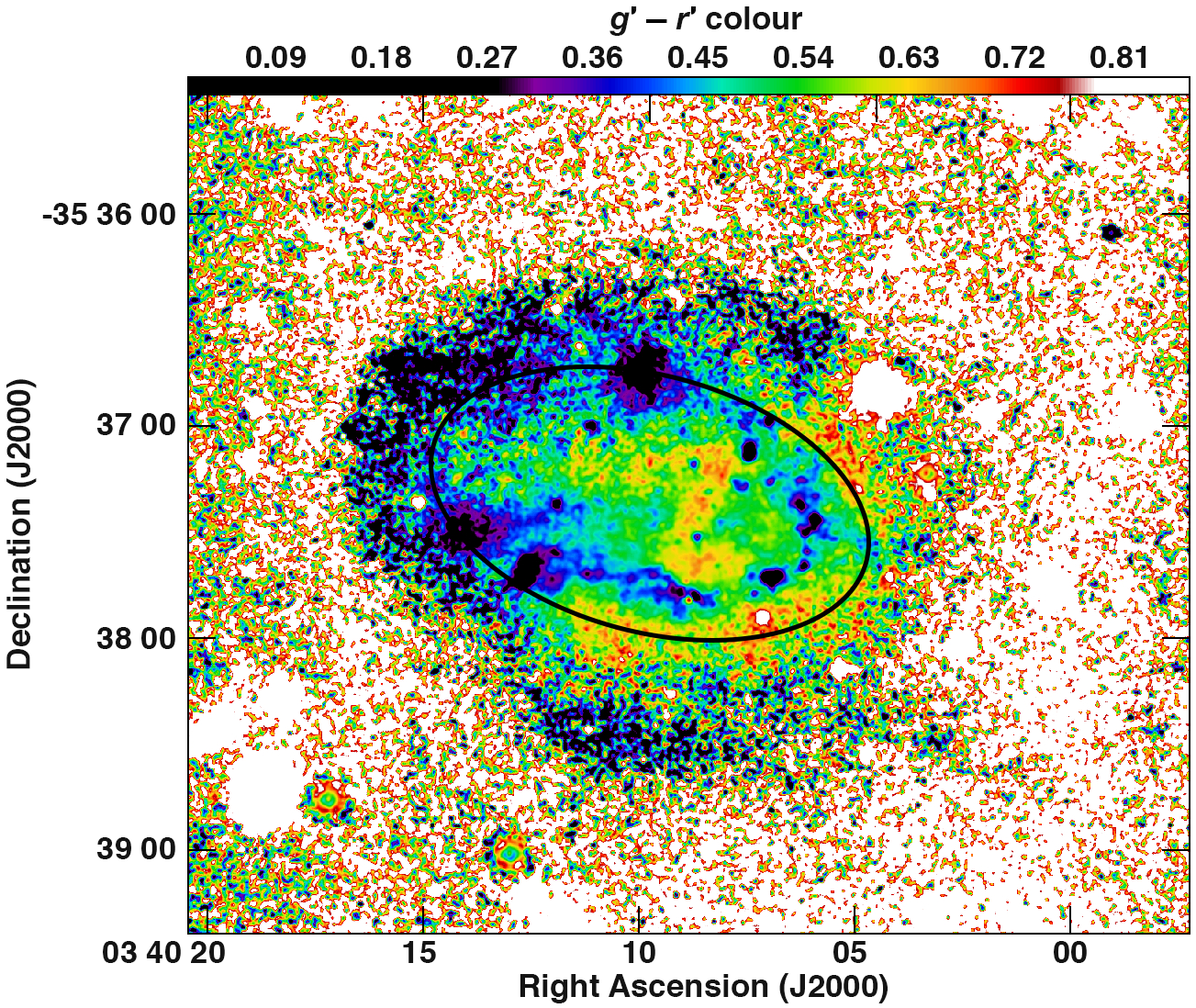}
  \caption{VST $g'-r'$ colour image of NGC~1427A.  The black ellipse is the same as the ellipses shown in Figs~\ref{fig:chan} and \ref{fig:ugri}.
\label{fig:blue_red}}
\end{center}
\end{figure}

As seen in Fig.~\ref{fig:ugri}, NGC~1427A has an underlying elliptical shape visible in all four bands.  The northern stellar clump is also detectable in the optical images.  We find, for the first time, an extended region of low surface brightness stellar light southwest of the star-forming region.  Furthermore, we also detect a stellar over-density `bump' extending toward the south, indicated in Fig.~\ref{fig:ugri}b.  Fig.~\ref{fig:blue_red} is a $g' - r'$ colour image of NGC 1427A from VST observations that further highlights the irregular optical morphology of NGC~1427A.  We describe the key stellar features in the following subsections.

\subsubsection{Stellar core}
\label{subsec:stellar_core}

The stellar core region is the bright central area of NGC~1427A, represented in Figs~\ref{fig:chan}, \ref{fig:ugri} and \ref{fig:blue_red} as a 125 $\times$ 70 arcsec ellipse with a position angle of 75 degree centred at 03h 40m 09.7s, -35\textdegree~37$'$ 23$''$.  As previously reported in the literature, there is active star-formation along its southwest portion, which appears brightest in the $g'$-band.  Fig.~\ref{fig:blue_red} shows an irregular mixing of blue and red colours as there is a red arch external (and southwest) to the blue star-forming region and then a `z' pattern of red stars near the centre.  Most of the bright H{\sc{i}} emission detected by ATCA coincides with this stellar core area (see Fig.~\ref{fig:chan}). 

\subsubsection{Northern stellar clump}
\label{subsec:nsc_optical}

The northern stellar clump is located at 03h 40m 10s, -35\textdegree~36$'$ 45$''$ and as shown in Fig.~\ref{fig:blue_red}, is one of the bluest regions in NGC~1427A.  Brightest in the $g'$-band, this clump is quite distinct from the other star-forming regions in the stellar core.  The deep optical images are able to resolve some of the internal structure of this stellar clump and there appears to be hints of a slight arrow-shape pointing north with an extremely faint stellar trail to the south/southwest (see Fig.~\ref{fig:ugri}b), which is also detectable in the optical imaging by \citet{hil1997}.  Using $H\alpha$ spectroscopy observations, \citet{cha2000} measure the heliocentric velocity of this feature to be $\sim$2029 km s$^{-1}$ and well within the H{\sc{i}} velocity range of NGC~1427A.

\subsubsection{Southwest extension}
\label{subsec:sw_optical}

There is a substantial amount of diffuse light extending towards the southwest of NGC~1427A.  This feature is $\sim$30 mag arcsec$^{-2}$ and in the redder bands extends over half a disk-length away from the stellar core.  There is a foreground star to the west of NGC~1427A (located at 03h 41m 45s, -35\textdegree~28$'$ 57$''$); however, the reflection halo of that star appears to have negligible contribution to the stellar light in the southwest extension.  This extension is located southwest of the bright star-forming region in the stellar core.  Such extended emission would not be expected if ram pressure forces impacted that side of the galaxy, which will be further discuss in Section~\ref{sec:discuss}.

\subsubsection{Southern bump}
\label{subsec:sb_optical}

The deep VST optical images also reveal a stellar over-density bump extending toward the south, indicated by the arrow in Fig.~\ref{fig:ugri}b.  This bump is less extended but somewhat brighter in the $g'$- and $r'$-bands than the southwest extension and is similar in colour as the northeast portion of NGC~1427A (see Fig.~\ref{fig:blue_red}).  The southern bump appears to spatially coincides with the base of the H{\sc{i}} tail of NGC~1427A.  In fact, while the VST images are able to detect fairly diffuse and extended features, there appears to be no substantial stellar counterpart for the H{\sc{i}} tail of NGC~1427A.
 
\section{Discussion}
\label{sec:discuss}

NGC~1427A has long been regarded as a good candidate for studying the effects of ram pressure stripping in the Fornax cluster.  A number of previous studies concluded that the disturbed optical appearance of NGC~1427A is due to its passage through and interaction with the dense ICM (e.g.~\citealt{cha2000}; \citealt{mor2015}).  Our resolved H{\sc{i}} observations and deep optical imaging of NGC~1427A allow us to study the distribution of gas and stars in the outskirts of this galaxy and constrain the processes affecting its irregular morphology.

\subsection{Is NGC 1427A located within the Fornax Cluster?}
\label{sec:distance}

Definitive distance measurements of NGC 1427A and its position within the Fornax cluster could determine which dynamical processes are affecting this object.  If the galaxy is located in the outskirts, away from the cluster potential, then tidal interactions would clearly give rise to its irregular optical morphology.  If it is deep within the cluster, then ram pressure would be the cause of this dIrr's arrow-shaped appearance.  

Optical spectroscopy by \citet{dri2001a} show that a significant number of star-forming dwarf galaxies are infalling onto the Fornax cluster.  Based on its position and recessional velocity, NGC~1427A could be part of this infalling population.  A comparison between the turnovers of the globular cluster luminosity function (GCLF) of NGC~1427A and NGC 1399 (the central cluster galaxy) places the former 3.2 Mpc $\pm$ 2.5 Mpc (statistic) $\pm$ 1.6 Mpc (systematic) in front of the latter \citep{geo2006}.  NGC~1427A might therefore be several Mpc away from the cluster centre, where ram pressure would be ineffective \citep{vol2001}.  However, the large statistic and systematic error on this estimate makes it difficult to reach a definitive verdict.  There is still the possibility that NGC~1427A is located within a region where interactions with the ICM can be a factor.

The most prominent spatially (in projection) and spectrally close neighbour to NGC~1427A is NGC~1404, a giant elliptical with an x-ray envelope that is currently being distorted by its infall towards NGC~1399 \citep{jon1997}. Distance measurements using a wide variety of indicators suggest that NGC~1404 is either $\sim$1 Mpc behind (e.g.~\citealt{fer2000}; \citealt{ton2001}; \citealt{jen2003}) or over 2 Mpc in front of (e.g.~\citealt{liu2002}; \citealt{tul2013}) other major cluster members.  Considering the uncertainty on the only independent distance measurement for NGC~1427A (i.e.~\citealt{geo2006}) and the variance in methods used to estimate distances to different types of galaxies, it is difficult to analyze any relationship (or lack thereof) between NGC~1427A and NGC~1404.  

Overall, due to the uncertainty in previously obtained distance measurements as well as the high velocity dispersion in the cluster environment that prohibits using radial velocity measurements as a proxy for distance, it is quite challenging to determine whether NGC~1427A is located within the Fornax cluster.  Accordingly, we discuss and compare our observed results in the context of ram pressure, tidal interactions, and a combination of these two processes in the following sections.

\subsection{Ram pressure hypothesis}
\label{sec:RPH}

Under the ram pressure hypothesis and given the location of its younger blue stars (as shown in Fig.~\ref{fig:ugri}a), NGC~1427A would be moving in a southwest direction within the cluster.  The ram pressure wind would therefore be acting towards the northeast.  Ram pressure is particularly effective at displacing gas at the outskirts of a galaxy relative to its stellar body (e.g.~\citealt{chu2009}; \citealt{ken2014}; \citealt{ken2015}; \citealt{abr2016} ).  For example, ESO137-001 is an infalling galaxy in the Norma cluster that has an extended gaseous tail and a modestly intact interior stellar region \citep{sun2010}. 

If ram pressure is acting on NGC~1427A, any H{\sc{i}} at large galactic radii should form a tail extending towards the northeast, relative to the centre of the galaxy.  However, our detection of a $>$20 kpc long H{\sc{i}} tail pointing perpendicular to this direction (i.e.~towards the southeast) contradicts the ram pressure hypothesis.  Additionally, the diffuse stellar extension as well as the faint H{\sc{i}} feature extending towards the southwest of the galaxy greatly opposes the expected ram pressure induced compression along that region.  

\citet{jon1997} and \citet{pao2002} do show that the x-ray envelope of NGC~1399 extends out, in projection, to NGC~1427A.  However, given the current uncertainty on the position of NGC~1427A relative to the Fornax cluster volume, this dIrr is not necessarily immersed within this hot gas and the evidence for the interaction between this galaxy and the ICM remains poor.  We note that NGC~1427A appears to be an x-ray source in those images, which could be interpreted as another result against ram pressure.  Presumably, if ram pressure was occurring, it would have removed the hot x-ray halo prior to stripping the cold H{\sc{i}} disk of this galaxy.

\subsection{Tidal interactions}
\label{sec:tidal}

If the arrow-shaped appearance of NGC~1427A is not the result of ram pressure, then the other obvious explanation would be that tidal interactions have caused the recent burst in star-formation.  The most conspicuous interloper would be the northern stellar clump of NGC~1427A.  Previous work ruled out this stellar clump as an intruder based on kinematical alignment \citep{cha2000}; nevertheless, it has been shown that some H{\sc{i}} satellites, which are accreted by larger galaxies, join the rotation of the parent's disk \citep{san2008}. This northern clump is also very blue, even compared to the other star-forming regions, which could indicate that it is still actively interacting with NGC~1427A.

The overall stellar component of NGC~1427A appears to be redder at larger radii and extends well beyond the galaxy's star-forming region, implying that star-formation is being triggered within the disk.  Additionally, the irregular mixing of colours in the stellar core possibly indicates the presence of two objects in the process of merging.  Depending on the initial parameters of the interaction, it is quite likely that the northern stellar clump started as a separate object and is the main contributor to the optical appearance of NGC~1427A.  

Alternatively, we explore the possibility that the tidal disturbance was triggered by a recent fly-by of another galaxy in the cluster.  The nearest HIPASS detected H{\sc{i}}-rich galaxies to NGC~1427A are 
ESO358-051 (HIPASS~J0341--34; 03h 41m 06s, -35\textdegree~56$'$ 02$''$, $F_{H_I}$ = 4.84 Jy km s$^{-1}$, $v_{H_I}$ = 1734 km s$^{-1}$),
ESO358-060 (HIPASS~J0345--35; 03h 45m 12s, -35\textdegree~34$'$ 07$''$, $F_{H_I}$ = 11.34 Jy km s$^{-1}$, $v_{H_I}$ = 803 km s$^{-1}$, and
NGC~1437A (HIPASS~J0342--36; 03h 42m 52s, -35\textdegree~17$'$ 26$''$, $F_{H_I}$ = 7.61 Jy km s$^{-1}$, $v_{H_I}$ = 895 km s$^{-1}$; \citealt{wau2002}).
The ATCA data for these galaxies detect comparable amounts of H{\sc{i}} to the values measured from HIPASS.  

ESO358-051 is located $>$200 kpc and $>$200 km s$^{-1}$ from NGC~1427A, while ESO358-060 is $\sim$350 kpc and $>$1200 km s$^{-1}$ away.  Both these galaxies show no clear morphological signs of being involved in a recent fly-by interaction.  Whereas, NGC~1437A has a similar arrow-shaped optical appearance as NGC~1427A and seems to be travelling in a southeast direction (based on the location of its own star-forming region) that is parallel with the orientation of NGC~1427A's H{\sc{i}} tail.  NGC~1437A has about one third the H{\sc{i}} mass as NGC~1427A and the velocity difference between these dIrrs is $\sim$1150 km s$^{-1}$, which is three times higher than the velocity dispersion of the Fornax cluster.  However, if these two galaxies experienced a previous flyby interaction and their velocity difference translates into their movement away from one another, then it would take a few hundred Myr to move 300 kpc (i.e. their projected separation distance) apart, which is considerably longer than the recent starburst episode occurring 4 Myr ago as measured by \citet{mor2015}.  The ATCA observations also detect unresolved H{\sc{i}} centred on NGC~1437; however, this galaxy has a fairly regular optical appearance that, similar to the previously discussed ESO galaxies, has no indication of recent tidal interactions.  

Overall, the ATCA data rule out signs of tidal disturbances with a projected linear size of $\sim$6 kpc (at a distance of 20 Mpc) down to a column density of $3\times10^{19}$ atoms cm${-2}$; however, with a 1.4 $\times$ 1 arcmin beam, several H{\sc{i}} detections are only marginally resolved.  In the currently available catalogues of Fornax region galaxies with known redshifts (i.e.~\citealt{dri2001b}; \citealt{mor2007}; \citealt{bla2009}), there is also no clearly apparent external tidal disturber of NGC~1427A.  The deep VST observations do detect a multitude of low surface brightness galaxies, two of which are -- in projection -- located near the tip of H{\sc{i}} tail \citep{ven2017}; however, without reliable distance measurements it is difficult to establish any physical association. 

\subsection{Tidal interactions + ram pressure}
\label{sec:tidal+RP}

We have effectively ruled out ram pressure acting from the southwest as the cause of the arrow-shaped optical morphology of NGC~1427A.  It seems most-likely that the northern stellar clump was once a separate object that is now merging with the main body of NGC~1427A.  This tidal interaction could also be responsible for the formation the newly detect H{\sc{i}} tail.  Although tidal tails typically form in pairs and generally have associated stellar components (\citealt{too1972}; \citealt{kav2012}), it is possible to have single tails and to form tidal debris with no detectable optical counterpart (e.g.~\citealt{chu2007}; \citealt{lee2014}).  We note that there is a hint of a counter-tail extending towards the north of the H{\sc{i}} core, however, this feature is only marginally detected (see Section~\ref{subsec:HI_core}).

The H{\sc{i}} tail could have formed as it is currently seen or, dependent on the location of NGC~1427A within the Fornax cluster, ram pressure could have a role in shaping this tail.  Tidal interactions could have initially expelled H{\sc{i}} to the galaxy's outskirts.  Subsequently, ram pressure could have swept together any tidally formed H{\sc{i}} tails creating a single trailing tail (see \citealt{chu2007}).  This sequence of events would be consistent with the presence of the southern stellar bump, which spatially coincides with the base of the H{\sc{i}} tail.  Starting from such bump, the H{\sc{i}} could have been moved to larger radii by ram pressure, leaving the stellar body relatively intact.   In this case, the implied movement of NGC~1427A within the cluster is towards the northwest, in the direction of the cluster centre.  Such motion would indicate a more radial orbit of NGC~1427A within Fornax than previously suggested by the literature.  




\section{Conclusions}
\label{sec:conclude}

We have detected an H{\sc{i}} tail and extended stellar emission that shed new light on the recent history of NGC~1427A.  The spatial position and distance of this dIrr indicates a possible association with the Fornax cluster \citep{dri2001b}; however, without accurate distance measurements, it is difficult to determine its exact location relative to the cluster and its current direction of travel.  Our new data rule out a ram pressure origin for the arrow-shaped optical appearance of NGC~1427A.  There is significant evidence that suggests a previous tidal interaction (with another Fornax cluster member) or a recent merging event has occurred to induce the recent starburst episode.  The irregularly mixed optical colours in the core of NGC~1427A and its distinct northern stellar clump favour the latter scenario.

This same interaction event could have formed the H{\sc{i}} tail in situ.  Alternatively, a combination of tidal and ram pressure forces could have swept any tidally formed tails into the currently detectable H{\sc{i}} tail, thereby indicating that NGC~1427A is travelling in a northwest direction, towards the cluster centre.  
Regardless of the exact combination of mechanisms that are affecting NGC~1427A, our new observations do make it apparent that tidal interactions are predominantly at play in this system.

\section*{Acknowledgements}

We thank the anonymous reviewer for his/her detailed suggestions and insightful comments to improve the clarity of this paper.  The Australia Telescope Compact Array is part of the Australia Telescope National Facility which is funded by the Australian Government for operation as a National Facility managed by CSIRO.  This project has received funding from the European Research Council (ERC) under the European Union's Horizon 2020 research and innovation programme (grant agreement no. 679627; project name FORNAX).  BC is the recipient of an Australian Research Council Future Fellowship (FT120100660).





\begin{thebibliography}{99}
\bibitem[\protect\citeauthoryear{Abramson et al.}{2016}]{abr2016}
Abramson A., Kenney J., Crowl H., Tal T., 2016, AJ, 152, 32
\bibitem[\protect\citeauthoryear{Bertin \& Arnouts}{1996}]{ber1996}
Bertin E., Arnouts S., 1996, A\&AS, 117, 393
\bibitem[\protect\citeauthoryear{Blakeslee et al.}{2009}]{bla2009}
Blakeslee J.P., et al., 2009, ApJ, 694, 556 
\bibitem[\protect\citeauthoryear{Bournaud et al.}{2007}]{bou2007}
Bournaud F., et al., 2007, Science, 316, 1166 
\bibitem[\protect\citeauthoryear{Bureau et al.}{1996}]{bur1996}
Bureau M, Mould J.R., Staveley-Smith L., 1996, ApJ, 463, 60 
\bibitem[\protect\citeauthoryear{Chaname et al.}{2000}]{cha2000}
Chaname J., Infante L., Reisenegger A., 2000, ApJ, 530, 96 
\bibitem[\protect\citeauthoryear{Chung et al.}{2007}]{chu2007}
Chung A., van Gorkom J.H., Kenney J., Vollmer B., 2007, ApJ, 659, 115
\bibitem[\protect\citeauthoryear{Chung et al.}{2009}]{chu2009}
Chung A., van Gorkom J.H., Kenney J., Crowl H., Vollmer B., 2009, AJ, 138, 1741
\bibitem[\protect\citeauthoryear{Cellone \& Forte}{1997}]{cel1997}
Cellone S.A., Forte J.C., 1997, AJ, 113, 1239
\bibitem[\protect\citeauthoryear{D'Abrusco et al.}{2016}]{dab2016}
D'Abrusco R., et al., 2016, ApJ, 819, 31 
\bibitem[\protect\citeauthoryear{Drinkwater et al.}{2001a}]{dri2001a}
Drinkwater M.J., Gregg M.D., Colless M., 2001a, ApJ, 548, L139  
\bibitem[\protect\citeauthoryear{Drinkwater et al.}{2001b}]{dri2001b}
Drinkwater M.J., Gregg M.D., Holman B.A., Brown M.J.I., 2001b, MNRAS, 326, 1076
\bibitem[\protect\citeauthoryear{Ferrarese et al.}{2000}]{fer2000}
Ferrarese L., et al., 2000, ApJ, 529, 745 
\bibitem[\protect\citeauthoryear{Georgiev et al.}{2006}]{geo2006}
Georgiev I.Y., Hilker M., Puzia T.H., Chaname J., Mieske S., Goudfrooij P., Reisenegger A., Infante L., 2006, A\&A, 452, 141
\bibitem[\protect\citeauthoryear{Gunn \& Gott}{1972}]{gun1972}
Gunn J.E., Gott J.R., 1972, ApJ, 176, 1
\bibitem[\protect\citeauthoryear{Hilker et al.}{1997}]{hil1997}
Hilker M., Bomans D.J., Infante L., Kissler-Patig M., 1997, A\&A, 327, 562
\bibitem[\protect\citeauthoryear{Iodice et al.}{2016}]{iod2016}
Iodice E., at al., 2016, ApJ, 820, 42 
\bibitem[\protect\citeauthoryear{Iodice et al.}{2017}]{iod2017}
Iodice E., at al., 2017, ApJ (submitted) 
\bibitem[\protect\citeauthoryear{Jensen et al.}{2003}]{jen2003}
Jensen J., Tonry J., Barris B., Thompson R., Liu M., Rieke M., Ajhar E., Blakeslee J., 2003, ApJ, 583, 712
\bibitem[\protect\citeauthoryear{Jones et al.}{1997}]{jon1997}
Jones C., Stern C., Forman W., Breen J., David L., Tucker W., 1997, ApJ, 482, 143
\bibitem[\protect\citeauthoryear{Kaviraj et al.}{2012}]{kav2012}
Kaviraj S., Darg D., Lintott C., Schawinski K., Silk J., 2012, MNRAS, 419, 70
\bibitem[\protect\citeauthoryear{Kenney et al.}{2014}]{ken2014}
Kenney J.D.P., Geha M., Jachym P., Crowl H.H., Dague W., Chung A., van Gorkom J., Vollmer B., 2014, ApJ, 780, 119
\bibitem[\protect\citeauthoryear{Kenney, Abramson \& Bravo-Alfaro}{2015}]{ken2015}
Kenney J.D.P., Abramson A., Bravo-Alfaro H., 2015, AJ, 150, 59
\bibitem[\protect\citeauthoryear{Koribalski et al.}{2004}]{kor2004}
Koribalski B.S., et al., 2004, AJ, 128, 16 
\bibitem[\protect\citeauthoryear{Lee-Waddell et al.}{2014}]{lee2014}
Lee-Waddell K., et al., 2014, MNRAS, 443, 3601
\bibitem[\protect\citeauthoryear{Liu, Graham \& Charlot}{2002}]{liu2002}
Liu M., Graham J., Charlot S., 2002, ApJ, 564, 216 
\bibitem[\protect\citeauthoryear{McFarland et al.}{2013}]{mcf2013}
McFarland J.P., Verdoes-Kleijn G., Sikkema G., Helmich E.M., Boxhoorn D.R., Valentijn E.A., 2013, ExA, 35, 45
\bibitem[\protect\citeauthoryear{McMullin et al.}{2007}]{mcm2007}
McMullin J., Waters B., Schiebel D., Young W., Golap K., 2007, ASPC, 376, 127
\bibitem[\protect\citeauthoryear{Mihos et al.}{2005}]{mih2005}
Mihos J.C., Harding P., Feldmeier J., Morrison H., 2005, ApJ, 631, 41
\bibitem[\protect\citeauthoryear{Mora et al.}{2015}]{mor2015}
Mora M.D., Chaname J., Puzia T.H., 2015, AJ, 150, 93 
\bibitem[\protect\citeauthoryear{Morris et al.}{2007}]{mor2007}
Morris R.A.H., Phillipps S., Jones J.B., Drinkwater M.J., Gregg M.D., Couch W.J., Parker Q.A., Smith R.M., 2007 A\&A, 476, 59
\bibitem[\protect\citeauthoryear{Paolillo et al.}{2002}]{pao2002}
Paolillo M., Fabbiano G., Peres G., Kim D.-W., 2002, ApJ, 565, 883
\bibitem[\protect\citeauthoryear{Sancisi}{2008}]{san2008}
Sancisi R., Fraternali F., Oosterloo T., van der Hulst T., 2008, A\&ARv, 15, 189
\bibitem[\protect\citeauthoryear{Sault, Teuben \& Wright}{1995}]{sau1995}
Sault R. J., Teuben P. J., Wright M. C. H., 1995, in Shaw R. A., Payne H. E., Hayes J. J. E., eds, ASP Conf. Ser. Vol. 77, Astronomical Data Analysis Software and Systems IV. Astron. Soc. Pac., San Francisco, p. 433
\bibitem[\protect\citeauthoryear{Serra et al.}{2015}]{ser2015}
Serra P., 2015, MNRAS, 448, 1922 
\bibitem[\protect\citeauthoryear{Sun et al.}{2010}]{sun2010}
Sun M., Donahue M., Roediger E., Nulsen P.E.J., Voit G.M., Sarazin C., Forman W., Jones C., 2010, ApJ, 708, 946
\bibitem[\protect\citeauthoryear{Taylor et al.}{2011}]{tay2011}
Taylor E., et al., 2011, MNRAS, 418, 1587 
\bibitem[\protect\citeauthoryear{Toloba et al.}{2014}]{tol2014}
Toloba E., 2014, ApJ, 783, 120 
\bibitem[\protect\citeauthoryear{Tonry et al.}{2001}]{ton2001}
Tonry J., Dressler A., Blakeslee J., Ajhar E., Fletcher A., Luppino G., Metzger M., Moore C., 2001,ApJ, 546, 681
\bibitem[\protect\citeauthoryear{Toomre \& Toomre}{1972}]{too1972}
Toomre A., Toomre J., 1972, ApJ, 178, 623
\bibitem[\protect\citeauthoryear{Tully et al.}{2013}]{tul2013}
Tully R., et al., 2013, AJ, 146, 86 
\bibitem[\protect\citeauthoryear{Venhola et al.}{2017}]{ven2017}
Venhola A., et al., 2017, A\&A (accepted for publication) 
\bibitem[\protect\citeauthoryear{Vollmer et al.}{2001}]{vol2001}
Vollmer B., Cayatte V., Balkowski C., Duschl W. J., 2001, ApJ, 561, 708
\bibitem[\protect\citeauthoryear{Waugh et al.}{2002}]{wau2002}
Waugh M., et al. 2002, MNRAS, 337, 641 
\end{thebibliography}



\bsp	
\label{lastpage}
\end{document}